\documentclass[manuscript,screen]{acmart}

\usepackage{threeparttable}
\usepackage{multirow}
\usepackage{subcaption}
\usepackage{pdfpages}
\AtBeginDocument{%
  }

\setcopyright{acmlicensed}
\copyrightyear{2018}
\acmYear{2018}
\acmDOI{XXXXXXX.XXXXXXX}
\acmConference[Conference acronym 'XX]{Make sure to enter the correct
  conference title from your rights confirmation email}{June 03--05,
  2018}{Woodstock, NY}
\acmISBN{978-1-4503-XXXX-X/2018/06}

\begin{document}

\title{Seeing to Think? How Source Transparency Design Shapes Interactive Information Seeking and Evaluation in Conversational AI}

\author{Jiangen He}
\orcid{0000-0002-3950-6098}
\affiliation{%
  \institution{The University of Tennessee}
  \streetaddress{1345 Circle Park Drive}
  \city{Knoxville}
  \state{TN}
  \country{USA}}
\email{jiangen@utk.edu}

\author{Jiqun Liu}
%\authornote{Corresponding author.}
\orcid{0000-0003-3643-2182}
\affiliation{%
  \institution{The University of Oklahoma}
  \streetaddress{401 W Brooks Street}
  \city{Norman}
 \state{OK}
  \country{USA}}
\email{jiqunliu@ou.edu}

\renewcommand{\shortauthors}{Trovato et al.}

\begin{abstract}
Conversational AI systems increasingly function as primary interfaces for information seeking, yet how they present sources to support information evaluation remains under-explored. This paper investigates how source transparency design shapes interactive information seeking, trust, and critical engagement. We conducted a controlled between-subjects experiment (N=372) comparing four source presentation interfaces—\textsc{Collapsible}, \textsc{Hover Card}, \textsc{Footer}, and \textsc{Aligned Sidebar}—varying in visibility and accessibility. Using fine-grained behavioral analysis and automated critical thinking assessment, we found that interface design fundamentally alters exploration strategies and evidence integration. While the \textsc{Hover Card} interface facilitated seamless, on-demand verification during the task, the \textsc{Aligned Sidebar} uniquely mitigated the negative effects of information overload: as citation density increased, Sidebar users demonstrated significantly higher critical thinking and synthesis scores compared to other conditions. Our results highlight a trade-off between designs that support workflow fluency and those that enforce reflective verification, offering practical implications for designing adaptive and responsible conversational AI that fosters critical engagement with AI generated content.
\end{abstract}


\begin{CCSXML}
<ccs2012>
   <concept>
       <concept_id>10003120.10003121.10011748</concept_id>
       <concept_desc>Human-centered computing~Empirical studies in HCI</concept_desc>
       <concept_significance>500</concept_significance>
       </concept>
   <concept>
       <concept_id>10003120.10003121.10003122.10003334</concept_id>
       <concept_desc>Human-centered computing~User studies</concept_desc>
       <concept_significance>500</concept_significance>
       </concept>
   <concept>
       <concept_id>10003120.10003123.10011759</concept_id>
       <concept_desc>Human-centered computing~Empirical studies in interaction design</concept_desc>
       <concept_significance>300</concept_significance>
       </concept>

 </ccs2012>
\end{CCSXML}

\ccsdesc[500]{Human-centered computing~Empirical studies in HCI}
\ccsdesc[500]{Human-centered computing~User studies}
\ccsdesc[300]{Human-centered computing~Empirical studies in interaction design}

\keywords{Conversational AI, Source Transparency, Human-AI Interaction, Information Seeking, Information Evaluation, Critical Thinking, Interface Design, Large Language Models, User Study}
  %\caption{Seattle Mariners at Spring Training, 2010.}
 % \Description{Enjoying the baseball game from the third-base
 % seats. Ichiro Suzuki preparing to bat.}
 % \label{fig:teaser}

\received{20 February 2007}
\received[revised]{12 March 2009}
\received[accepted]{5 June 2009}

\maketitle

\section{Introduction}

Conversational AI systems are increasingly used as primary interfaces for information seeking, learning, and evidence-based writing. Unlike traditional search engines, these systems actively frame and summarize information through dialogue, shaping how users interpret evidence, assess sources, and decide when further verification is needed. This shift raises important concerns about how users form trust, how they recognize uncertainty and biases, and how they integrate multiple sources into their reasoning and task completion~\cite{liu2023toward}. Prior research shows that interface transparency can strongly influence users’ mental models and reliance on algorithmic outputs~\cite{kizilcec2016much, eslami2015always}, while interaction design choices affect whether users critically reflect on system suggestions or accept them with minimal scrutiny~\cite{amershi2019guidelines, binns2018s}. At the same time, studies of human–AI collaboration suggest that users often over-rely on fluent AI responses unless the interface encourages deliberate engagement~\cite{buccinca2021trust, ribeiro2016should}. Together, these findings highlight the importance of understanding interactive information seeking and evaluation as a core component of responsible conversational AI~\cite{ribeiro2016should}.

However, studying this process in conversational AI is particularly challenging because source use is mediated by interface design rather than direct document interaction. Source transparency is not only about whether citations exist, but also about how visible, accessible, and cognitively affordable they are during interaction. Research on algorithmic awareness shows that when system mechanisms are hidden, users may construct inaccurate explanations of how results are generated~\cite{eslami2015always, corbett2023interrogating}. Meanwhile, work on explainability and transparency demonstrates that explanations and documentation can fail to improve understanding, or even create false confidence or mistrust, depending on how they are presented~\cite{bordt2022post, mitchell2019model, liu2025boundedly}. In conversational AI systems, fluent language further blurs the boundary between generated content and external evidence, making it harder for users to distinguish, verify, and reconcile sources. As a result, citation interfaces may either support careful evaluation or subtly discourage it, depending on how they distribute attention and effort across reading, verifying, and writing activities~\cite{buccinca2021trust, vasconcelos2023explanations}.

Motivated by these challenges, this work investigates how \textit{source transparency design} shapes interactive information seeking and evaluation in conversational AI. We examine how different source display and access mechanisms (e.g. click, hover, listing) influence users’ exploration behavior, chat perception, and evidence use during writing tasks. Our research questions focus on the interaction between interface design and information density, and on how this interaction affects critical engagement with information sources cited by system responses. The novelty of this work lies in treating source transparency as a behavioral and interactional phenomenon rather than a purely normative design principle~\cite{mitchell2019model, turri2024transparency}. This study bridges research on human–AI interaction, algorithmic transparency, and critical reliance by grounding these discussions in fine-grained behavioral evidence~\cite{amershi2019guidelines, buccinca2021trust, bordt2022post}. In addition, the findings will offer concrete guidance and practical implications for designing citation and source presentation interfaces that support verification without overwhelming users and faciltiate evidence-based critical thinking and learning, contributing to more trustworthy and responsible conversational AI experiences~\cite{ribeiro2016should, vasconcelos2023explanations, liu2025report, li2023trustworthy}. 

In this paper, we investigate the impact of citation presentation on user behavior and trust in AI-generated text. We guide our study with the following \textit{Research Questions} (RQs):

\begin{description}
\item[RQ1:] How do variations in source visibility and accessibility shape users' information-seeking strategies and integration workflows?
\item[RQ2:] To what extent does source presentation design influence perceived transparency and trust?
\item[RQ3:] How does the interplay between source presentation design and information density impact users' critical evaluation and synthesis of evidence in information evaluation?
\end{description}

This work makes several contributions to understanding source transparency in conversational AI. In addition, it frames source transparency as an interactional and behavioral phenomenon rather than a static documentation feature. Through a controlled interface experiment and fine-grained behavioral analysis, we provide empirical evidence on how variations in source saliency, accessibility, and information density shape users’ information-seeking strategies, trust perceptions, and evidence integration workflows. Besides advancing theoretical insight into conversational information interaction, this study also demonstrates how interface design can differentially support or hinder critical engagement under high information density in conversational information seeking. These findings offer practical guidance for designing transparent, trustworthy, and cognitively supportive conversational AI systems, and contribute to broader efforts to promote responsible and human-centered AI deployment and evaluation.

\section{Related Work}
Building on our framing of source transparency as an interactional and behavioral phenomenon in conversational AI, we situate this work within prior research on conversational information seeking, human–AI transparency, and interface-mediated information use.
\subsection{Conversational Information Seeking and Evaluation}
Conversational AI systems have transformed information seeking from a document-centered activity into a dialogic, natural-language-driven, and system-mediated process~\cite{yuan2025query}. Instead of navigating ranked lists, users increasingly rely on AI agents to synthesize, prioritize, and narrate information through natural language interaction~\cite{lajewska2024explainability}. Prior work in human–AI interaction shows that such mediation shapes users’ judgments, confidence, and verification behaviors~\cite{amershi2019guidelines, buccinca2021trust}. From an interactive information retrieval perspective, information seeking is inherently iterative and situated~\cite{liu2019task}, and conversational interfaces further blur the boundaries between retrieval, interpretation, and judgment~\cite{zamani2023conversational, ruthven2008interactive}. Generative responses compress multiple sources into a single narrative form, which can obscure uncertainty and disagreement while increasing perceived coherence and fluency~\cite{vasconcelos2023explanations, jakesch2023co}. As a result, conversational information seeking introduces new challenges for how users recognize uncertainty, compare evidence, and decide when additional exploration is necessary, particularly in tasks that require critical evaluation rather than simple fact lookup~\cite{turri2024transparency, corbett2023interrogating, gao2020recent, mo2025survey}.

\textit{Evaluation} in conversational information seeking therefore extends beyond factual correctness to include judgments of relevance, credibility, sufficiency, and argumentative support across sources~\cite{dalton2022conversational}. Research on boundedly rational search and AI-assisted decision making shows that users rely heavily on heuristics and interface cues under cognitive constraints~\cite{kiesel2021meant, azzopardi2021cognitive}, often defaulting to system suggestions unless interaction design introduces cognitive forcing or reflective friction~\cite{buccinca2021trust, shah2024envisioning, liu2025boundedly, chen2023reference}. At the same time, studies of algorithmic awareness indicate that users frequently hold incomplete or inaccurate mental models of how conversational systems retrieve and generate information~\cite{eslami2015always, kizilcec2016much, lee2019webuildai}, which further complicates their ability to critically evaluate AI-mediated evidence~\cite{singh2025enhancing}. Work on transparency and explanation mechanisms suggests that providing more information does not necessarily improve understanding, and can sometimes increase misplaced confidence or shallow verification~\cite{mitchell2019model, bordt2022post, he2025not}. While prior studies have examined trust, transparency, and reliance in conversational and generative systems~\cite{ribeiro2016should, corbett2023interrogating}, fewer have investigated how users behaviorally explore, revisit, and integrate sources during extended complex conversational tasks such as writing and argument construction. This gap motivates our focus on conversational information seeking and evaluation as an AI-faciliated interactional and interface-dependent process rather than a purely cognitive or perceptual outcome.

\subsection{Transparency and Explanation in Human-AI Interaction}
Transparency and explanation have been widely advocated as essential mechanisms for fostering appropriate trust, accountability, and user understanding in AI systems~\cite{Eyert2023Transparency, turri2024transparency}. Early work in human–computer interaction demonstrated that explanations influence how users interpret system behavior, diagnose errors, and form expectations about future performance~\cite{lim2009and, poursabzi2021manipulating}. Subsequent research emphasized that explanations are not neutral technical artifacts, but interactional constructs whose effectiveness depends on user goals, task context, and representational form~\cite{hohman2018visual, wang2019designing}. Users often prefer explanations that are concise, contrastive, and actionable rather than exhaustive, uncertain, or mathematically precise~\cite{liao2020questioning, al2024user}. At the same time, empirical studies show that explanations can alter users’ confidence and reliance patterns, sometimes increasing trust even when system outputs are incorrect or uncertain~\cite{Bansal2021Whole, lee2021included}. These findings challenge the implicit assumption that more explanation necessarily leads to better understanding, and instead suggest that transparency must be evaluated in terms of how it shapes human judgment and behavior.

Building on this perspective, more recent scholarship frames transparency as a situated and socio-technical process rather than a static property of AI models~\cite{ehsan2021expanding, el2024transparent, kim2023help}. Research on human-centered explainable AI shows that users actively question, reinterpret, and negotiate explanations based on their evolving mental models and task demands~\cite{liao2020questioning, Ehsan2020HumanCenteredXAI}. However, explanations can also anchor user interpretations, bias attention, and reinforce confirmation effects, particularly when explanations are presented as authoritative or definitive~\cite{Du2019InterpretableML, Gilpin2018ExplainingExplanations, liu2023behavioral}. Work on uncertainty visualization and interpretability evaluation further demonstrates that transparency cues may overwhelm users or redirect attention away from substantive evidence evaluation~\cite{Gilpin2018ExplainingExplanations}. While emerging studies have begun to examine explanation design in conversational and interactive AI systems~\cite{AlAnsari2024XAIReview, Zhang2025HumanCenteredXAI}, much of the literature still evaluates explanations in isolation from broader interaction workflows. Consequently, less is known about how transparency mechanisms interact with interface layout, information density, and task context to shape users’ real information evaluation activities~\cite{li2025matching}. This gap motivates our treatment of transparency not merely as an explanatory feature, but as an interactional design choice that fundamentally structures how users engage with and utilize AI-generated contents.

\subsection{Interface Design in User Interaction with Generated Contents}
Interface design fundamentally shapes how users perceive, navigate, and integrate generated information~\cite{norval2022disclosure, kim2024m}. Prior work in human–computer interaction has long shown that information interaction is driven not only by content quality, but by spatial organization, visual encoding, and interaction cost~\cite{kittur2008harnessing, marchionini2006exploratory, goyal2016effects}. In sensemaking and multi-document tasks, interface structures influence attention allocation and cognitive load, exploration breadth, and synthesis depth~\cite{jansen2017integrative, azzopardi2018measuring, schmitt2024role}. These effects become more pronounced when users must coordinate reading, note-taking, and writing under cognitive constraints, where even small interaction frictions can alter engagement patterns~\cite{reza2025co, packer2023designing, zhao2025making}. Studies further show that fragmented layouts and split attention reduce comprehension and integration, even when all relevant information is technically available~\cite{mcclure2009you, lee2024one, min2025malleable}. These studies jointly establish interface design as a core determinant of information behavior rather than a secondary usability concern in human-AI interaction.

In AI-mediated and generative systems, interface design becomes even more consequential because users interact with synthesized content that already carries an aura of authority~\cite{narayanan2025search, wolfe2024impact}. Research on AI-assisted writing, ideation, and programming demonstrates that interface affordances shape how users accept, verify, and reinterpret generated content, as well as how they attribute responsibility and authorship~\cite{bibi2024role, wang2024investigating}. In conversational and generative retrieval settings, scholars have emphasized the need for interaction designs that support comparison, grounding, and iterative refinement rather than one-shot acceptance~\cite{radlinski2017theoretical, trippas2020towards}. At the same time, interface designs that prioritize fluency and convenience may inadvertently discourage reflection, while overly dense displays can overwhelm users and reduce engagement. Despite these insights, existing work has rarely examined how interface layouts and access mechanisms shape users’ behavioral transitions among exploration, reading, citation, and writing when interacting with source-backed generated content~\cite{cuddihy2012effect, vannostrand2024actionable}. This gap is particularly salient in conversational AI, where generated responses, sources, and writing activities are tightly interwoven within a single workflow. Building on conversational information seeking and transparency research, our study positions interface design as a central mechanism through which conversational AI systems structure users’ information evaluation, evidence integration, and critical engagement.

%\subsection{Summary}
\section{Methods}

\subsection{Interface Design}

\begin{figure}[h]
    \centering
    \includegraphics[width=\textwidth]{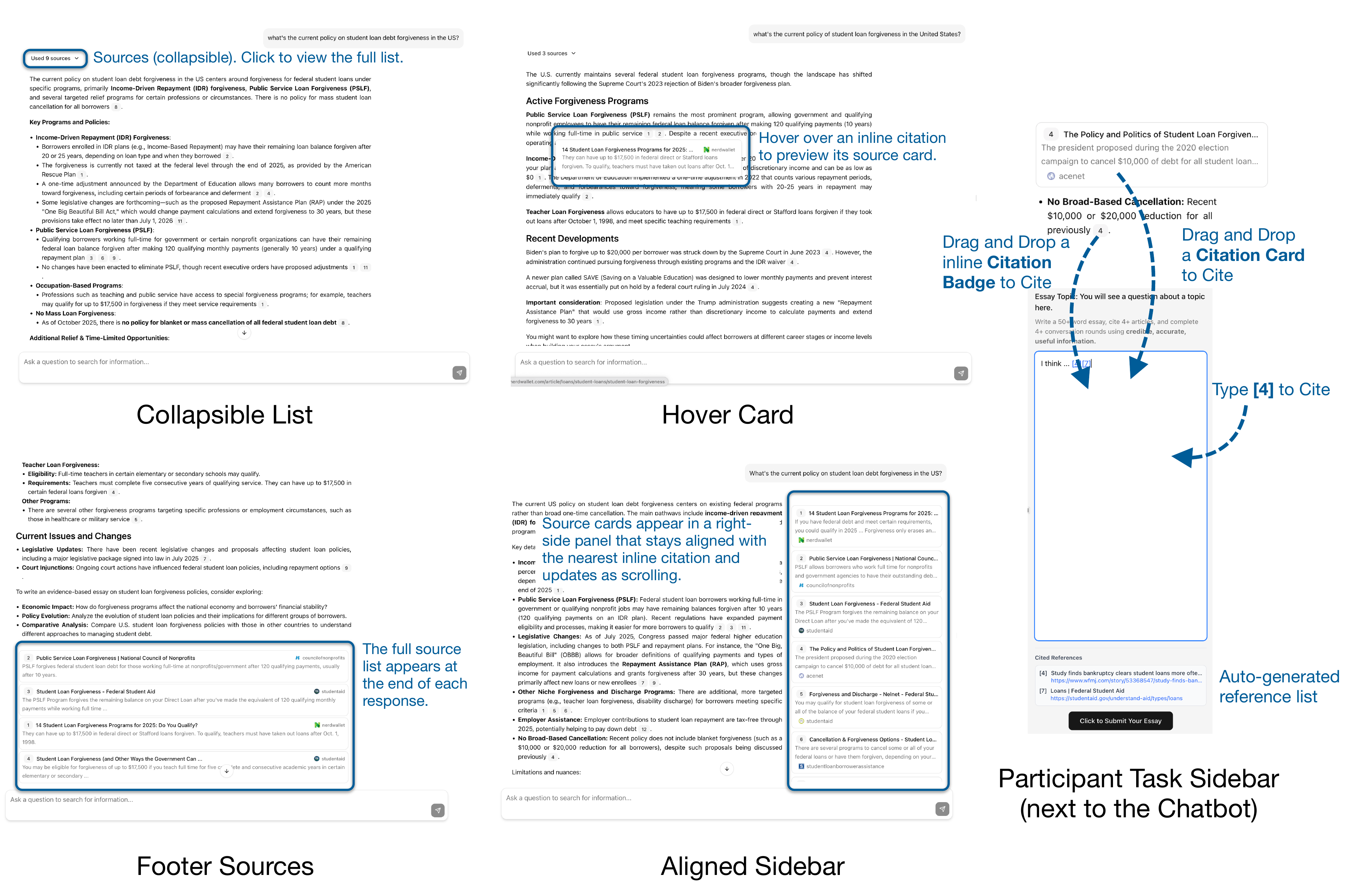}
    \caption{Interface conditions}
    \label{fig:interfaces}
\end{figure}

To investigate how different source presentation formats influence user experience and chat experience, we implemented four distinct interface conditions (see Figure~\ref{fig:interfaces}).
Except the baseline interface, the four interfaces with source presentation presents the same source details (title, domain name with favicon, and text snippet), but each interface varies in how sources are displayed and accessed. Across all four conditions, participants can read  sources by clicking on their badges or cards. They can also cite sources in their essay by dragging and dropping them directly into the essay editor to automatically insert a citation.

\subsubsection{\textsc{Collapsible} Interface}

In the collapsible interface condition, source badges appear inline within the text as clickable links.
A collapsible ``Used Sources'' button displays above the message content. Participants can toggle this button to reveal the complete list of source cards.

\subsubsection{\textsc{Hover Card} Interface}

In the hover card interface condition, when users hover over a citation badge, a card appears displaying the source details.
This design provides on-demand access to citation details, allowing users to quickly preview sources while maintaining their position in the text.

\subsubsection{\textsc{Footer} Interface}

In the footer interface condition, all source cards are grouped at the end of each response message. All source cards are displayed by default without the need to expand, compared to the collapsible interface. This approach mimics bibliography formatting, consolidating all reference information in a location. 

\subsubsection{\textsc{Aligned Sidebar} Interface}

The aligned sidebar interface condition presents source cards in a persistent sidebar adjacent to the main content.
Source cards are dynamically positioned to align vertically with their corresponding in-text citations, entering with a slide-in animation.
The interface employs the bounded isotonic regression (PAVA) to compute optimal non-overlapping positions for the cards. The display of source card updates dynamically in response to scrolling and window resizing.
Crucially, the layout algorithm implements a viewport-aware optimization during scrolling, dynamically adjusting card weights to prioritize the visibility of cards near the viewport center.
When citations appear multiple times within the text, positioning is weighted toward the first occurrence using exponential decay to prioritize initial encounters. The algorithm details is in Appendix Section \ref{sec:sidebar_algorithm}

We use the four interface conditions to manipulate the source visibility and accessibility. Visibility refers to the explicit display of source details, while accessibility refers to the proximity of source details to the in-text citations.
Based on these dimensions, the \textsc{Collapsible List} condition represents low visibility and low accessibility. The \textsc{Footer} condition provides high visibility but low accessibility. The \textsc{Hover Card} condition offers low visibility but high accessibility. Finally, the \textsc{Aligned Sidebar} condition achieves both high visibility and high accessibility.

\subsubsection{Citing Sources}
Citations could be integrated into the essay via two methods: a \textit{drag-and-drop} mechanism, allowing users to drag citation badges or cards directly into the editor, or a \textit{smart-typing} feature, where typing a citation index (e.g., ``[1]'') automatically converted the text into a linked citation. Once inserted, citations appeared in the editor as clickable, blue, bracketed links (e.g., [1]). Users could remove citations using standard text deletion methods (Backspace key), which treated the citation as a single unit. To assist with bibliography management, the system dynamically generated and updated a ``Cited References'' list below the editor, displaying the full details (title and URL) of all unique sources currently cited in the text. 

\subsection{AI Agent}
The AI agent was powered by Perplexity AI's \texttt{sonar-pro} model, configured with a temperature of 0.7.
The model utilized web search capabilities with medium search context size and user location set to `US' (United States) to retrieve real-time information from the internet and generate responses with inline citations. To ensure high-quality, evidentiary support, the agent operated under a system prompt (see Appendix~\ref{sec:system_prompt}) that defined its role as a research assistant. Key principles included minimizing bias, encouraging critical thinking, and supporting evidence-based writing without generating the essay itself. Responses were constrained to approximately 400 words.

\subsection{Procedure}

The study followed a between-subjects design with participants randomly assigned to one of the five interface conditions and one of the topics. The experimental workflow consisted of three phases: pre-survey, instruction, task, and post-survey.

At first, participants completed a pre-survey to gather demographic information and assess their prior knowledge, interest, and agreement with the policy topic (see Appendix~\ref{sec:survey_instruments}).
Second, Participants were directed to an instructional page featuring a tutorial video specific to their assigned interface condition (see links in Appendix~\ref{sec:instruction_videos}). The video demonstrated the core features of the system, specifically focusing on how to access and verify source information using the assigned citations display (e.g., hovering over cards in the \textsc{Hover} condition, or using the sidebar in the \textsc{Sidebar} condition). The system enforced a mandatory viewing period; the button to proceed to the main task remained disabled until the video concluded.
After the video, participants were transitioned to the main experimental environment, which featured a split-pane layout: a conversational AI interface occupied the left 75\% of the screen, while a text editor occupied the right 25\%. Each participant could see the assigned one of three policy topics and instructed to write an argumentative essay using information gathered from the AI. We chose argumentative essay writing as a naturalistic proxy for studying critical information use: unlike simple fact-finding, it requires multi-source synthesis, evidence weighing, and perspective integration~\cite{10.1145/3613904.3642459}. Critically, the writing task enables fine-grained capture of citation behavior—how users access, verify, and integrate sources—while producing a measurable artifact for assessing critical thinking outcomes. The specific topics and their assigned questions are detailed in Table~\ref{tab:task_topics}.

\begin{table}[h]
  \caption{Task Topics and Assigned Questions}
  \label{tab:task_topics}
  \small
  \renewcommand{\arraystretch}{1.2}
  \begin{tabular}{ll}
    \toprule
    \textbf{Topic} & \textbf{Assigned Question} \\
    \midrule
    Healthcare & Should the U.S. Government Provide Universal Health Care? \\
    News & Should the U.S. government impose stricter regulations on social media platforms to curb misinformation? \\
    Science & Should the U.S. permit clinical use of germline gene editing in humans? \\
    \bottomrule
  \end{tabular}
\end{table}
To guarantee sufficient interaction with the system, the task enforced three completion criteria: participants had to (1) write a minimum of 50 words, (2) complete at least four conversation rounds with the AI, and (3) cite at least four unique sources. Upon meeting all task requirements, participants submitted their essays and were automatically redirected to a Qualtrics post-study survey.

\paragraph{Data Collection}
During the session, the system captured granular behavioral metrics, including keystrokes, mouse movements, and interaction events (such as clicks and hovers) with precise timestamps, as well as AI responses and behavioral data on citing and writing. To ensure data quality, attention-check questions were embedded in both the pre- and post-study surveys. Participants who failed the pre-survey check were asked to return the study, while those who failed the post-survey check were excluded from the final analysis.

\subsection{Critical Thinking Assessment}

To examine how source presentation formats influence participants' critical engagement with information---a key component of critical thinking---we employed an LLM-based automated essay scoring framework~\cite{peczuh2025llmsupportedautomatedassessmentcritical}. This method offers a scalable and cost-effective method for quantifying fine-grained critical thinking subskills with reliability comparable to human experts, particularly for the information-centric skills relevant to our study. The assessment covers three broad skills and six specific subskills (Table~\ref{tab:ct_skills}). Notably, the subskills of \textit{Synthesizing Multiple Sources}, \textit{Evaluating Evidence Strength}, and \textit{Using Facts and Opinions} are directly relevant to analyzing how users integrate and evaluate external information.
\begin{table}[h]
  \caption{Critical Thinking Skills and Subskills Definitions}
  \label{tab:ct_skills}
  \small
  \renewcommand{\arraystretch}{1.25}
  \setlength{\tabcolsep}{3pt}
  \begin{tabular}{p{0.20\columnwidth}p{0.26\columnwidth}p{0.43\columnwidth}}
    \toprule
    \textbf{Skill} & \textbf{Subskill} & \textbf{Definition} \\
    \midrule
    \multirow{2}{=}{Information Analysis} & \textbf{Synthesizing} Multiple Sources & Effectively synthesizes multiple pieces of information \\
     & \textbf{Evaluating} Evidence Strength & Evaluates the strength and relevance of evidence used to form a conclusion \\
    \midrule
    \multirow{2}{=}{Argument Generation} & Using \textbf{Counterarguments} & Effectively addresses counterarguments \\
     & Using \textbf{Facts} and Opinions & Relies on data and/or facts over opinions \\
    \midrule
    \multirow{2}{=}{Logical Reasoning for Critical Thinking} & Drawing \textbf{Conclusions} & Draws specific conclusions \\
     & Using Logical \textbf{Fallacies} & Recognizes and avoids logical fallacies \\
    \bottomrule
  \end{tabular}
\end{table}

\section{Results}
In this section, we present our findings organized by our three research questions. First, we examine how interface designs influenced users' behavioral strategies and workflows (RQ1). Next, we analyze the effects on perceived transparency, trust, and user experience (RQ2). Finally, we investigate how these designs interacted with information density to shape critical thinking and evidence integration (RQ3).

\subsection{Participants}
We recruited a total of 372 participants via Prolific. All included participants passed the embedded attention checks and provided valid submission data. To ensure demographic diversity, we utilized Prolific's representative sampling feature, focusing on a balance across Political Affiliation and Ethnicity. The sample was stratified based on Sex, Age, Ethnicity (using Simplified US Census categories), and Political Affiliation. The final sample consisted of 190 females (50.9\%) and 183 males (49.1\%), with a mean age of 44.88 ($SD=15.30$). Participant ethnicity was diverse: 59.2\% White, 12.3\% Black, 11.5\% Mixed, 7.5\% Asian, and 9.4\% other. The median completion time was 22.88 minutes. Each participant was compensated \$4 for their participation. The study has been reviewed and approved by an Institutional Review Board.

\subsection{Interaction Analysis (RQ1)}
To analyze user behavior, we categorized actions into five distinct types: \textit{Explore} (interaction with citations), \textit{Engage} (reading), \textit{Cite} (adding citation), \textit{Write} (essay composition), and \textit{Prompt} (prompting AI).

Table~\ref{tab:behavior_metrics} presents the descriptive statistics for various behavioral metrics across the four interface conditions. We conducted Kruskal-Wallis H tests to identify significant differences.
When we analyzed user engagement levels (``Engage'') and interaction patterns (``Explore''), as expected, the \textsc{Collapsible} condition necessitated significantly more Badge Clicks ($Mean=4.03$) to reveal content ($p<.001$). In contrast, the \textsc{Sidebar} and \textsc{Footer} interfaces facilitated exploration through Card Hovers ($Mean=33.09$ and $26.04$, respectively), significantly exceeding the interaction rates of the \textsc{Hover} and \textsc{Collapsible} groups ($p<.001$).

Most interestingly, we observed distinct engagement strategies in reading behavior. Pairwise comparisons (Dunn's test) revealed that the \textsc{Collapsible} condition led to significantly more reading sessions ($Mean=4.48$, $p<.001$) compared to all other conditions, driving the highest total reading duration ($69.33$s). This suggests that the intentional act of expanding content commits users to more frequent reading bouts. In terms of sustained attention, the \textsc{Footer} condition achieved the highest \textit{average} reading duration per session ($11.65$s). Pairwise analysis confirmed that this duration was significantly longer than the \textsc{Hover} condition ($7.00$s, $p=.007$), though differences with \textsc{Collapsible} ($9.88$s, $p=.25$) and \textsc{Sidebar} ($8.86$s, $p=.16$) were not statistically significant. This indicates that while \textsc{Collapsible} encourages frequency, the \textsc{Footer} placement may better support longer, uninterrupted reading flows, avoiding the fragmented attention potentially induced by \textsc{Hover} interactions.

We also analyzed how participants incorporated citations into their essays. While the total number of citations remained consistent across groups, the method of citation varied significantly. Participants in the \textsc{Collapsible} and \textsc{Hover} conditions relied heavily on clicking badges to cite ($Mean=4.64$ and $5.19$, respectively), whereas those in the \textsc{Sidebar} and \textsc{Footer} conditions frequently used card clicks ($Mean=3.09$ and $2.52$). Interestingly, we found no significant difference in the number of citations added via typing ($p=0.344$), with participants across all conditions manually typing approximately 1.5 to 2 citations on average. This suggests that despite the availability of convenient ``drag-to-cite'' features in the context of AI responses, a portion of users consistently prefer or default to manual entry.

\begin{table*}[htbp]
\centering
\caption{Behavioral Metrics across Interface Conditions (Mean $\pm$ SD)}
\label{tab:behavior_metrics}
\small
\begin{tabular}{lcccccc}
\toprule
\textbf{Metric} & \textsc{Collapsible} (N=95) & \textsc{Hover} (N=91) & \textsc{Footer} (N=92) & \textsc{Sidebar} (N=93) & \textbf{H} & \textbf{p-value} \\
\midrule
\multicolumn{7}{l}{\textit{Prompt}} \\
Conversation Rounds & 4.82 (2.74) & 4.85 (1.34) & 4.79 (1.26) & 4.96 (1.76) & 2.34 & 0.506 \\
Msg Total Length & 370.05 (408.09) & 342.67 (200.56) & 380.92 (353.32) & 372.61 (244.84) & 0.53 & 0.911 \\
Msg Avg Length & 72.21 (31.17) & 69.98 (37.17) & 75.71 (49.47) & 74.93 (48.15) & 0.50 & 0.918 \\
Msg Typing & 103.00 (84.82) & 97.56 (61.89) & 105.12 (84.82) & 107.91 (86.97) & 0.41 & 0.938 \\

\midrule
\multicolumn{7}{l}{\textit{Explore}} \\
Card Clicks & 0.46 (1.56) & 0.00 (0.00) & 1.27 (2.46) & 1.85 (3.33) & 77.72 & <.001*** \\
Card Hovers & 5.47 (18.43) & 6.18 (4.84) & 26.04 (18.93) & 33.09 (23.97) & 211.93 & <.001*** \\
Badge Clicks & 4.03 (5.61) & 1.22 (2.14) & 0.97 (2.22) & 0.76 (1.66) & 45.24 & <.001*** \\
Badge Hovers & 18.19 (18.45) & 16.60 (12.01) & 10.06 (10.10) & 7.97 (8.29) & 54.24 & <.001*** \\
\midrule
\multicolumn{7}{l}{\textit{Engage}} \\
Reading Sessions & 4.48 (5.64) & 1.22 (2.14) & 2.23 (3.88) & 2.61 (3.66) & 25.67 & <.001*** \\
Reading Total Duration (s) & 69.33 (136.46) & 22.60 (60.19) & 40.95 (87.53) & 52.48 (140.73) & 20.71 & <.001*** \\
Reading Avg Duration (s) & 9.88 (13.78) & 7.00 (20.43) & 11.65 (22.08) & 8.86 (20.52) & 15.30 & 0.002** \\
\midrule
\multicolumn{7}{l}{\textit{Cite}} \\
Essay Citations & 5.42 (2.28) & 5.41 (1.84) & 5.50 (2.37) & 5.50 (2.05) & 1.34 & 0.719 \\
\hspace{3mm}By Badge & 4.64 (3.99) & 5.19 (3.80) & 3.36 (4.12) & 2.48 (2.71) & 34.13 & <.001*** \\
\hspace{3mm}By Card & 0.40 (1.40) & 0.25 (0.97) & 2.52 (3.12) & 3.09 (4.18) & 91.61 & <.001*** \\
\hspace{3mm}By Typing & 1.99 (3.36) & 1.65 (3.43) & 1.55 (3.34) & 1.74 (3.08) & 3.33 & 0.344 \\
Citations Deleted & 1.16 (2.40) & 0.83 (1.05) & 1.30 (2.53) & 1.28 (1.81) & 2.64 & 0.451 \\

\midrule
\multicolumn{7}{l}{\textit{Write}} \\
Essay Word Count & 135.32 (91.60) & 120.34 (66.16) & 124.09 (67.76) & 118.78 (53.17) & 1.77 & 0.620 \\
Focus Sessions & 11.78 (9.06) & 8.53 (5.76) & 11.09 (6.45) & 10.87 (7.17) & 14.27 & 0.003** \\
Essay Typing Time (s) & 304.14 (177.89) & 293.58 (174.11) & 305.49 (198.71) & 281.86 (150.59) & 0.85 & 0.837 \\

\midrule
\multicolumn{7}{l}{\textit{AI}} \\
AI Citations Shown & 26.45 (13.24) & 26.40 (7.66) & 25.68 (6.75) & 26.86 (12.02) & 1.75 & 0.627 \\
AI Msg Total Length & 8302 (6005) & 7719 (2358) & 7741 (2199) & 7637 (2883) & 2.06 & 0.560 \\
AI Avg Msg Length & 1629 (262) & 1600 (310) & 1626 (327) & 1536 (277) & 7.02 & 0.071 \\
\bottomrule
\end{tabular}
\begin{tablenotes}
\item \textit{Note:} * $p<0.05$, ** $p<0.01$, *** $p<0.001$. Kruskal-Wallis H Test.
\end{tablenotes}
\end{table*}

\subsection{Behavioral Patterns (RQ1)}

\subsubsection{Behavioral Changes across Messages}

\begin{figure}[h]
  \centering
  \includegraphics[width=1\linewidth]{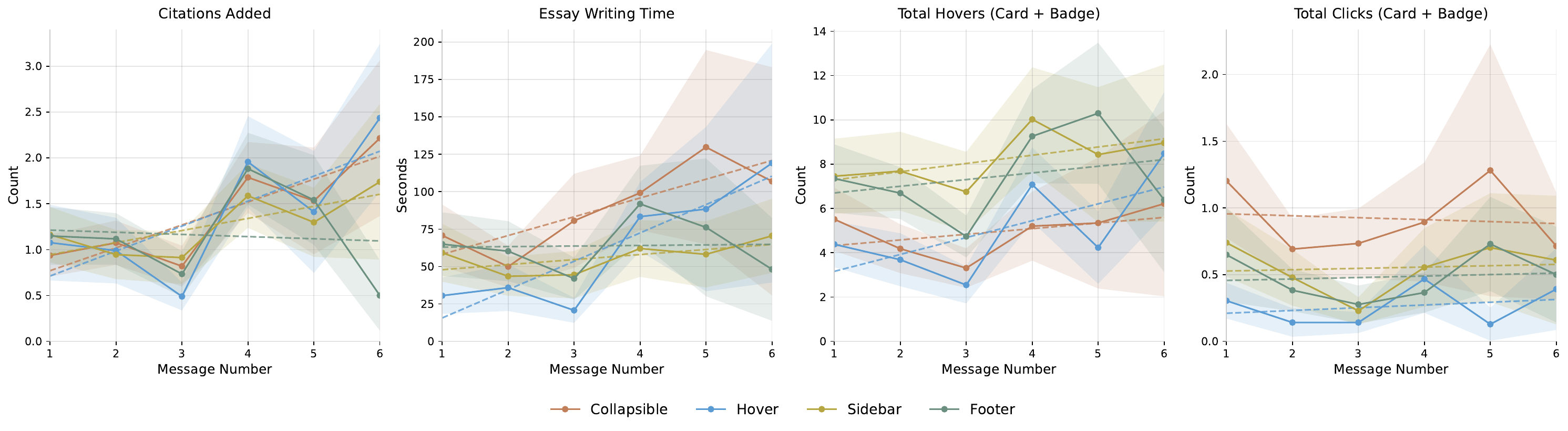}
  \caption{Behavioral metrics across message rounds (1-6). The charts display trends across the four interface conditions.}
  \label{fig:behavior_metrics_dashboard}
\end{figure}

Figure~\ref{fig:behavior_metrics_dashboard} visualizes the temporal evolution of key behavioral metrics over the course of the conversation. We observed participation attrition in the optional rounds: while all participants completed the mandatory four messages, 60.8\% of participants (N=226) chose to end the session immediately after the fourth message. Subsequently, 18.0\% (N=67) concluded after the fifth message, and 9.4\% (N=35) ended after the sixth message. The temporal changes reveal persistent differences in interaction styles between conditions. 

Citation patterns diverged significantly across conditions over time. Participants in the \textsc{Hover}, \textsc{Collapsible}, and \textsc{Sidebar}  condition demonstrated a positive trajectory in citing citations over messages. Conversely, usage in the \textsc{Footer} condition declined sharply after the forth message. Importantly, this drop was not due to general disengagement; analysis of writing time confirmed that \textsc{Footer} participants maintained consistent effort (avg. writing duration $\sim$60-90s) throughout the session. This dissociation suggests that the increasing interface friction of accessing citations buried in previous message footers actively discouraged their use, whereas the immediate, inline availability of \textsc{Hover} cards supported sustained and deepening integration of evidence.

Beyond citation counts, interaction patterns revealed a fundamental distinction between exploration and deep engagement. Across all conditions, participants exhibited a dominant ``content skimming'' strategy, heavily favoring hovering (Exploration) over clicking (Reading). While click rates for full-text verification remained consistently low ($<1$ per round) and uniform across groups, exploration trajectories differed by design. Notably, participants in the \textsc{Collapsible} condition failed to show increased hovering engagement as the conversation progressed. This suggests that interface interventions influence not only the immediate frequency of interaction but also the \textit{deepening} of engagement over time. Designs characterized by low visibility or accessibility, such as the \textsc{Collapsible} and \textsc{Footer} interfaces, appear to impose interaction costs that discourage users from developing deeper verification habits as the task evolves.

\subsubsection{Behavioral Sequence Analysis}

To understand how different source presentation designs influence user engagement and information seeking strategies, we conducted a behavior sequence analysis following the methodology of \cite{cole2015}. We modeled the sequence of these user actions as a Markov chain to calculate the probability of transitioning between states.

\paragraph{Dominant Interaction Loops.} Table~\ref{tab:sequence_transitions_top6} reveals the most frequent behavioral transitions for each condition. The self-loop \textit{Explore} $\to$ \textit{Explore} was the dominant behavior in all interfaces, reflecting the core task of scanning through citations. However, the intensity of this loop varied: \textit{Sidebar} (32.4\%, $P=0.615$) and \textit{Footer} (30.0\%, $P=0.604$) users spent significantly more of their total interaction budget on continuous exploration compared to \textit{Collapsible} users (19.2\%, $P=0.537$). This suggests that visible, list-based interfaces (Sidebar, Footer) encourage ``batch processing'' of citations, whereas the Collapsible design leads to more fragmented exploration.

\begin{figure}[h]
  \centering
  \includegraphics[width=0.75\linewidth]{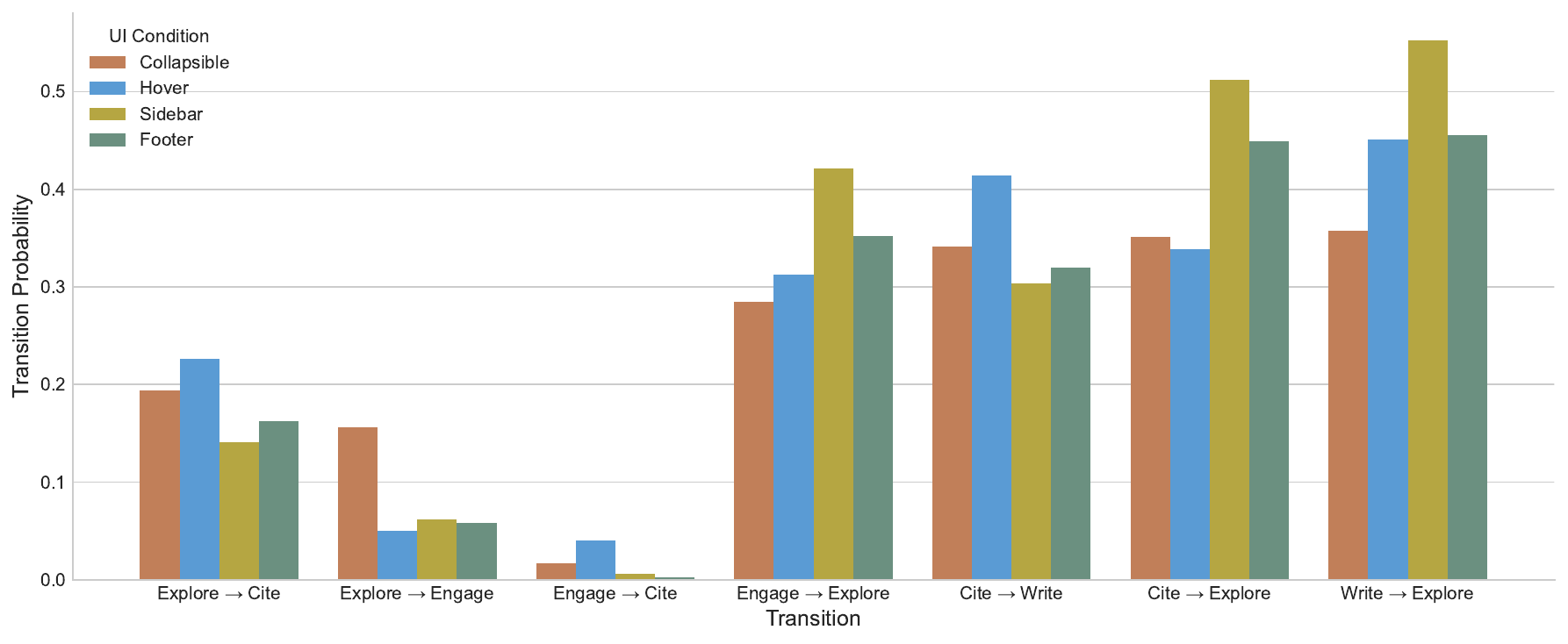}
  \caption{Visualizing high-probability transitions. \textsc{Hover} condition (top-right) shows strong links between Exploration, Citation, and Writing.}
  \label{fig:key_transitions}
\end{figure}

\paragraph{Writing Workflow Integration.} A critical differentiator found in the sequence analysis is the ``return to flow.'' As shown in Table~\ref{tab:sequence_transitions_top6} and Figure~\ref{fig:key_transitions}, the \textit{Hover} condition is the unique interface where the transition \textit{Cite} $\to$ \textit{Write} appears in the top six most frequent behaviors (5.8\% of all actions, with a conditional probability of $P=0.393$). This indicates that floating cards allow users to seamlessly verify and insert a citation, then immediately resume text generation. In contrast, \textit{Footer} and \textit{Sidebar} users frequently transitioned from \textit{Cite} back to \textit{Explore} (e.g., Footer: 5.4\%, $P=0.441$), suggesting that the list-based layout prompts users to continue searching or confirming other sources rather than returning to their writing task.

\paragraph{Verification and Deep Engagement.} The \textit{Sidebar} interface strongly supported verification during drafting, with \textit{Write} $\to$ \textit{Explore} emerging as its second most frequent transition (8.0\% frequency, $P=0.520$). This "check-as-you-write" behavior was less prominent in the \textit{Collapsible} condition (6.2\%, $P=0.316$). Interestingly, the \textit{Collapsible} condition was the only one where \textit{Engage} $\to$ \textit{Engage} (deep reading self-loop) appeared in the top behaviors (7.9\%), implying that when users did open the collapsible panel, they committed to reading the content more deeply than the scanning behaviors observed in other conditions.

\begin{table}[t]
  \centering
  \caption{Top 6 Most Frequent Behavior Transitions by Condition. Organized in two rows to compare interaction patterns: Collapsible and Hover (top) vs. Footer and Sidebar (bottom).}
  \label{tab:sequence_transitions_top6}
  \renewcommand{\arraystretch}{1.1}
  \small
  \begin{tabular}{llcc|llcc}
    \toprule
    \textbf{Transition} & \textbf{Freq} & \textbf{\%} & \textbf{Prob.} & \textbf{Transition} & \textbf{Freq} & \textbf{\%} & \textbf{Prob.} \\
    \midrule
    \multicolumn{4}{l|}{\textsc{Collapsible}} & \multicolumn{4}{l}{\textsc{Hover}} \\
    Exp $\to$ Exp & 1113 & 19.2\% & 0.537 & Exp $\to$ Exp & 1321 & 26.7\% & 0.623 \\
    Prompt $\to$ Prompt & 559 & 9.7\% & 0.616 & Prompt $\to$ Prompt & 581 & 11.7\% & 0.646 \\
    Engage $\to$ Engage & 456 & 7.9\% & 0.553 & Exp $\to$ Cite & 459 & 9.3\% & 0.217 \\
    Exp $\to$ Cite & 388 & 6.7\% & 0.187 & Write $\to$ Exp & 341 & 6.9\% & 0.387 \\
    Write $\to$ Write & 386 & 6.7\% & 0.339 & Cite $\to$ Write & 287 & 5.8\% & 0.393 \\
    Write $\to$ Exp & 360 & 6.2\% & 0.316 & Write $\to$ Write & 239 & 4.8\% & 0.271 \\
    \midrule
    \multicolumn{4}{l|}{\textsc{Footer}} & \multicolumn{4}{l}{\textsc{Sidebar}} \\
    Exp $\to$ Exp & 2017 & 30.0\% & 0.604 & Exp $\to$ Exp & 2330 & 32.4\% & 0.615 \\
    Exp $\to$ Cite & 498 & 7.4\% & 0.149 & Write $\to$ Exp & 576 & 8.0\% & 0.520 \\
    Write $\to$ Exp & 477 & 7.1\% & 0.419 & Prompt $\to$ Prompt & 536 & 7.5\% & 0.586 \\
    Prompt $\to$ Prompt & 464 & 6.9\% & 0.508 & Exp $\to$ Cite & 504 & 7.0\% & 0.133 \\
    Cite $\to$ Exp & 361 & 5.4\% & 0.441 & Exp $\to$ Write & 485 & 6.8\% & 0.128 \\
    Exp $\to$ Write & 355 & 5.3\% & 0.106 & Cite $\to$ Exp & 396 & 5.5\% & 0.501 \\
    \bottomrule
  \end{tabular}
\end{table}

\subsection{Perception (RQ2)}

Table~\ref{tab:perception_results} presents the results of the adjusted ordinal regression models for the perception and user experience outcomes.

\begin{table}[htbp]
    \centering
    \begin{threeparttable}
    \caption{Summary of Adjusted Regression Models for Perception and Experience Outcomes}
    \label{tab:perception_results}
    \small
    \setlength{\tabcolsep}{3pt}
    \begin{tabular}{lccccccc}
        \toprule
        \textbf{Predictor} & \textbf{Transparency} & \textbf{Trust} & \textbf{Confidence} & \textbf{Satisfaction} & \textbf{Usefulness} & \textbf{Quality} & \textbf{Workload} \\
        \midrule
        \multicolumn{8}{l}{\textit{UI Conditions (ref: \textsc{Collapsible})}} \\
        UI: \textsc{Sidebar}             & -0.37    & -0.44    & -0.62    & -1.26    & -1.41*   & -0.43    & -0.66    \\
        UI: \textsc{Hover}               & 1.15     & 0.24     & 0.29     & 0.25     & -0.18    & -0.13    & -1.12    \\
        UI: \textsc{Footer}              & 0.38     & -0.33    & 0.87     & 0.27     & -0.24    & -0.06    & -0.37    \\
        \midrule
        \multicolumn{8}{l}{\textit{Interactions}} \\
        \textsc{Sidebar} $\times$ Citations & 0.03     & 0.02     & 0.03     & 0.05     & 0.04*    & 0.01     & 0.03     \\
        \textsc{Hover} $\times$ Citations& -0.03    & -0.01    & 0.00     & -0.01    & -0.01    & 0.00     & 0.05*    \\
        \textsc{Footer} $\times$ Citations  & -0.01    & 0.02     & -0.03    & 0.00     & 0.01     & -0.00    & 0.02     \\
        % \midrule
        % \multicolumn{8}{l}{\textit{Controls}} \\
        % Age                     & -0.00    & -0.00    & -0.01    & 0.01     & 0.00     & -0.00    & 0.02**   \\
        % Sex Female              & 0.42*    & -0.03    & -0.16    & 0.24     & 0.19     & 0.47*    & 0.16     \\
        % Education               & 0.04     & -0.13    & -0.02    & 0.03     & 0.07     & 0.07     & 0.17     \\
        % Llm\_usage\_freq        & -0.06    & -0.19**  & -0.09    & -0.16*   & -0.19**  & -0.16*   & -0.10    \\
        % Search\_usage\_freq     & -0.17    & 0.28     & -0.07    & -0.11    & 0.20     & 0.22     & 0.05     \\
        % Eval\_rel\_confidence   & 0.97***  & 0.80***  & 0.88***  & 0.90***  & 0.78***  & 0.59**   & -0.16    \\
        % Eval\_trus\_confidence  & -0.06    & 0.01     & 0.11     & -0.15    & 0.18     & 0.17     & -0.06    \\
        % Knowledge\_pre          & -0.14    & -0.10    & 0.23**   & -0.08    & -0.10    & -0.03    & -0.12    \\
        % Interest\_pre           & 0.06     & -0.01    & 0.02     & 0.09     & 0.02     & 0.02     & 0.18*    \\
        % Agreement\_pre          & 0.11     & 0.16*    & 0.06     & -0.00    & 0.07     & 0.11     & -0.01    \\
        % Humility                & 0.41**   & 0.36**   & -0.03    & 0.24     & 0.32*    & 0.27     & 0.02     \\
        % \midrule
        % $R^2$ (Pseudo)          & 0.09     & 0.07     & 0.09     & 0.08     & 0.09     & 0.06     & 0.03     \\
        % N                       & 372      & 372      & 372      & 372      & 372      & 372      & 372      \\
        \bottomrule
    \end{tabular}
    \begin{tablenotes}
        \item * $p<0.05$, ** $p<0.01$, *** $p<0.001$. Complete model with controls for all predictors see Appendix Table~\ref{tab:perception_results_complete}.
    \end{tablenotes}
    \end{threeparttable}
\end{table}

We found that the \textsc{Sidebar} interface had a significant negative main effect on Usefulness ($\beta = -1.41, p < 0.05$), but showed a significant positive interaction with citation count ($\beta = 0.04, p = 0.05$). This suggests that while the \textsc{Sidebar} design was  perceived as less useful than the \textsc{Collapsible} baseline, its usefulness improved as the number of citations increased. It might be because the high saliency of source presentation in \textsc{Sidebar} made participants perceive the low density of information sources easier. 
Additionally, the \textsc{Hover} interface showed a significant positive interaction with citation count for Workload ($\beta = 0.05, p < 0.05$), indicating that the cognitive load of using the \textsc{Hover} interface increased with source density. We did not observe significant effects of interface conditions on Transparency, Trust, or Confidence in the regression models.

\subsection{Critical Evidence-based Writing (RQ3)}

\begin{figure*}[h]
  \centering
  \begin{subfigure}[b]{0.42\textwidth}
    \centering
    \includegraphics[width=\linewidth]{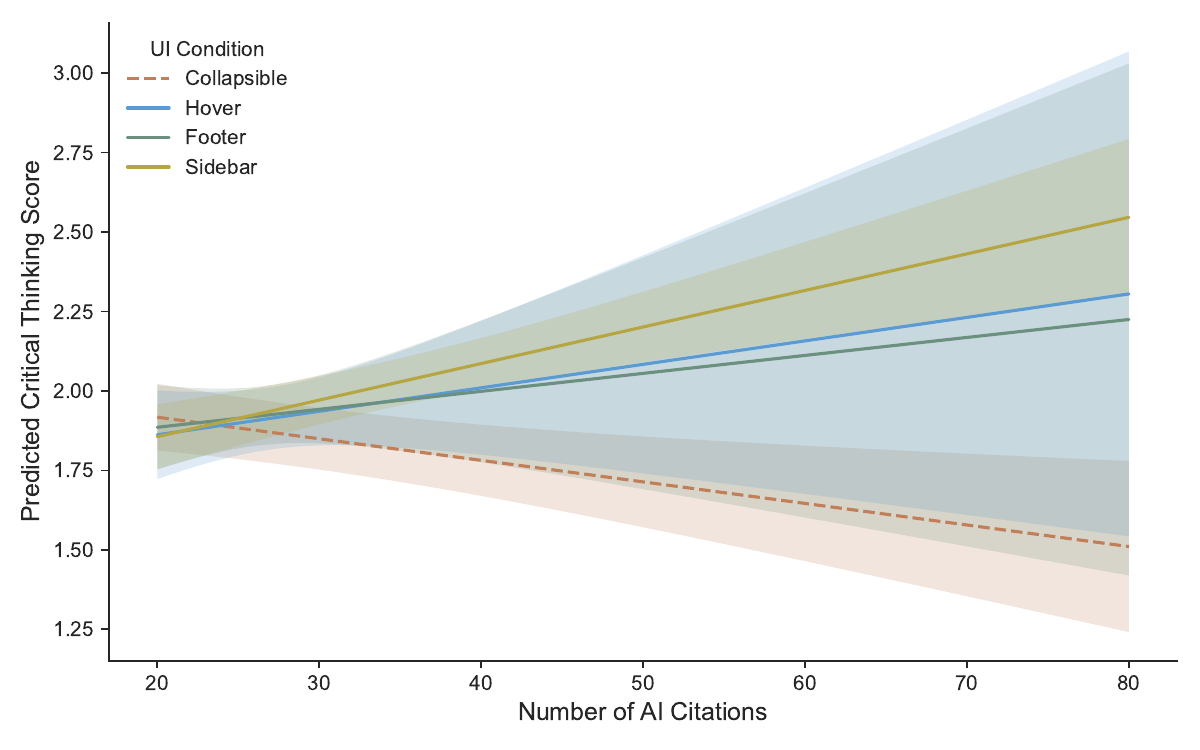}
    \caption{Overall Critical Thinking Score}
    \label{fig:interaction_overall}
  \end{subfigure}
  \hfill
  \begin{subfigure}[b]{0.42\textwidth}
    \centering
    \includegraphics[width=\linewidth]{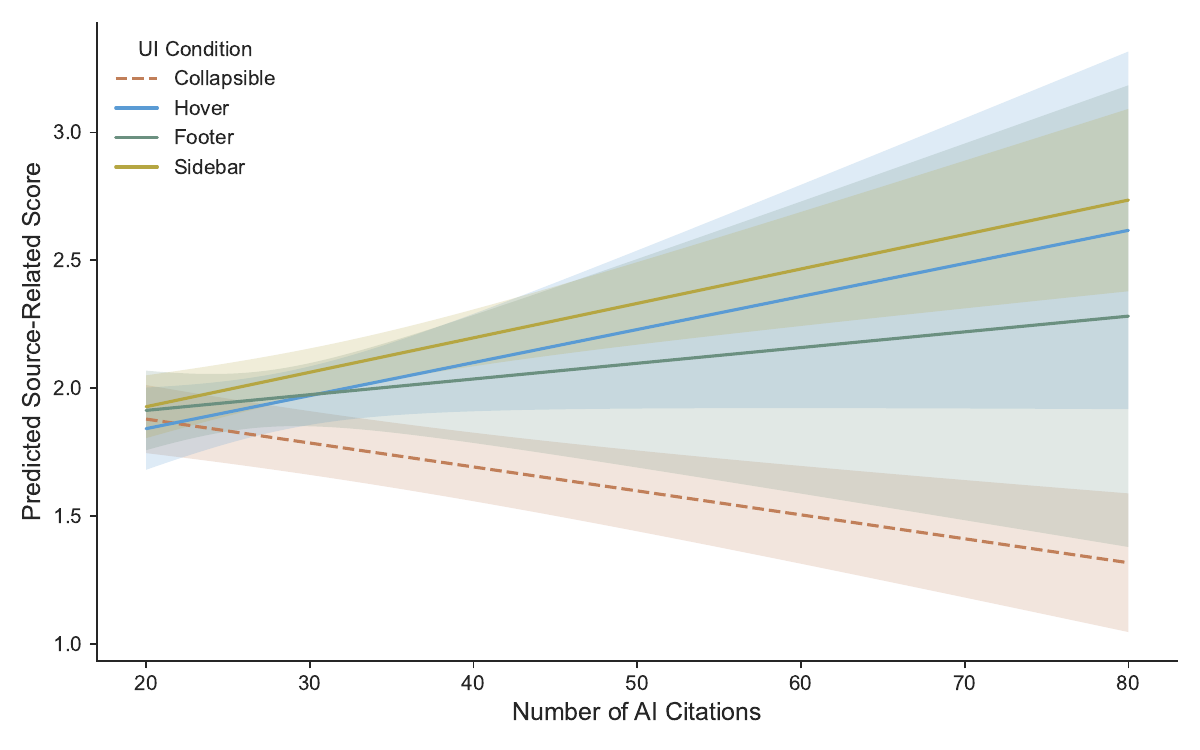}
    \caption{Source-Related Critical Thinking Score}
    \label{fig:interaction_source}
  \end{subfigure}
  \caption{Interaction effect of UI design and information density (number of citations) on critical thinking scores. Shaded areas represent 90\% confidence intervals.}
  \label{fig:interaction_plot}
\end{figure*}

The regression analysis revealed a significant interaction effect between UI design and information density (measured by the number of citations in AI responses). As shown in Figure~\ref{fig:interaction_plot}, while critical thinking scores generally remain flat or decline with increased citation volume for most designs, the \textsc{Sidebar} condition exhibits a positive trajectory. Specifically, as the number of AI citations increases, participants in the \textsc{Sidebar} group demonstrate higher critical thinking scores compared to the \textsc{Collapsible} (baseline), \textsc{Hover}, and \textsc{Footer} conditions. This suggests that the \textsc{Sidebar} interface may better support users in managing high information density, allowing them to maintain or even improve their critical engagement when presented with more sources.

\begin{table*}[htbp]
    \centering
    \begin{threeparttable}
    \caption{Summary of Adjusted Regression Models (Model 2) for All Critical Thinking Outcomes}
    \label{tab:model2_summary}
    \small
    \setlength{\tabcolsep}{3pt}
    \begin{tabular}{lcccccccc}
        \toprule
        \textbf{Predictor} & \textbf{Synthes.} & \textbf{Eval.} & \textbf{Counterarg.} & \textbf{Facts/Opinions} & \textbf{Concl.} & \textbf{Fallacies} & \textbf{Source-Rel.} & \textbf{Overall} \\
        \midrule
        \multicolumn{9}{l}{\textit{UI Conditions (ref: \textsc{Collapsible})}} \\
        UI: \textsc{Sidebar}             & -1.53*   & -0.55    & -0.86    & -1.09    & -1.05    & -1.78*   & -0.36    & -0.36*  \\
        UI: \textsc{Hover}               & -1.16    & -1.48    & -0.18    & -0.90    & -0.25    & -0.68    & -0.42    & -0.24   \\
        UI: \textsc{Footer}              & -0.95    & 0.15     & -0.45    & -0.76    & -1.52    & -0.72    & -0.27    & -0.25   \\
        Citations               & -0.02    & -0.01    & 0.00     & -0.02    & -0.02    & -0.04*   & -0.01    & -0.01   \\
        \midrule
        \multicolumn{9}{l}{\textit{Interactions}} \\
        \textsc{Sidebar} $\times$ Citations & 0.07**   & 0.05     & 0.02     & 0.07**   & 0.04     & 0.07**   & 0.02**   & 0.02**  \\
        \textsc{Hover} $\times$ Citations& 0.06     & 0.08*    & -0.00    & 0.05     & 0.02     & 0.04     & 0.02*    & 0.01    \\
        \textsc{Footer} $\times$ Citations  & 0.04     & 0.01     & 0.01     & 0.05     & 0.06     & 0.03     & 0.01     & 0.01    \\
        % \midrule
        % \multicolumn{9}{l}{\textit{Controls}} \\
        % Age                     & -0.03*** & -0.03*** & -0.01    & -0.03*** & -0.02**  & -0.03*** & -0.01*** & -0.01***\\
        % Sex Female              & 0.35     & 0.42     & 0.16     & 0.46*    & 0.14     & 0.47*    & 0.16*    & 0.12*   \\
        % Education               & 0.49***  & 0.28*    & 0.04     & 0.39***  & 0.16     & 0.36**   & 0.13***  & 0.09**  \\
        % Llm\_usage\_freq        & 0.09     & 0.01     & 0.04     & 0.07     & 0.04     & 0.02     & 0.02     & 0.02    \\
        % Search\_usage\_freq     & 0.06     & 0.08     & -0.36*   & -0.11    & -0.11    & -0.09    & 0.01     & -0.04   \\
        % Eval\_rel\_confidence   & 0.05     & 0.32     & 0.15     & 0.08     & 0.11     & 0.10     & 0.05     & 0.05    \\
        % Eval\_trus\_confidence  & 0.27     & 0.06     & 0.03     & 0.31     & 0.08     & 0.14     & 0.07     & 0.05    \\
        % Knowledge\_pre          & -0.18*   & -0.28**  & -0.22**  & -0.16    & -0.29**  & -0.21*   & -0.06*   & -0.07***\\
        % Interest\_pre           & 0.18*    & 0.16     & 0.04     & 0.06     & 0.09     & 0.12     & 0.03     & 0.03    \\
        % Agreement\_pre          & 0.08     & 0.18*    & 0.08     & 0.14     & 0.06     & 0.17*    & 0.05     & 0.04*   \\
        % Humility                & 0.19     & 0.16     & 0.15     & 0.16     & 0.22     & 0.52***  & 0.09     & 0.08*   \\
        % \midrule
        % $R^2$                   & 0.07     & 0.09     & 0.03     & 0.07     & 0.05     & 0.10     & 0.17     & 0.18    \\
        % N                       & 371      & 371      & 371      & 371      & 371      & 371      & 365      & 371     \\
        \bottomrule
    \end{tabular}
    \begin{tablenotes}
        \item * $p<0.05$, ** $p<0.01$, *** $p<0.001$. Complete model with controls for all predictors see Appendix Table~\ref{tab:model2_summary_complete}.
    \end{tablenotes}
    \end{threeparttable}
\end{table*}
This advantage was driven by significant improvements in specific subskills. Table~\ref{tab:model2_summary} shows that the \textsc{Sidebar} condition had significant positive interactions for \textit{Synthesizing} ($\beta=0.07, p<0.01$), \textit{Using Facts and Opinions} ($\beta=0.07, p<0.01$), \textit{Using Logical Fallacies} ($\beta=0.07, p<0.01$), and notably, \textit{Source-Related} critical thinking (a composite of Synthesis, Evaluation, and Facts; $\beta=0.02, p<0.01$). As visualized in Figure~\ref{fig:interaction_source}, while the \textsc{Sidebar} condition exhibits a positive trajectory in source-related scores as citation volume increases, other conditions flatline or decline. This indicates that the persistent visibility of sources in the \textsc{Sidebar} specifically aids users in \textit{connecting} and critically evaluating multiple sources when information density is high.

Interestingly, the \textsc{Sidebar} condition exhibited negative main effects for several outcomes (e.g., \textit{Synthesizing}: $\beta=-1.53, p<0.05$; \textit{Overall}: $\beta=-0.36, p<0.05$), implying that at low information density, the split-attention cost of the \textsc{Sidebar} might initially hamper performance. However, its strong positive interaction effects allow it to surpass other designs as information becomes more dense.

In contrast, the interaction landscape for \textsc{Hover} condition revealed a unique trade-off. While it did not match the \textsc{Sidebar} in synthesis, \textsc{Hover} showed a significant positive interaction specifically for \textit{Evaluating} ($\beta=0.08, p<0.05$). This suggests that the interface affordance of the \textsc{Hover} card—presenting a single source in isolation upon demand—facilitates a more targeted scrutiny of individual evidence reliability. Thus, while \textsc{Sidebar} supports the \textit{breadth} of critical thinking required for high-volume synthesis, \textsc{Hover} appears to enhance the \textit{depth} of source evaluation.

\section{Discussion}
Our results demonstrate that source transparency is fundamentally contextual and intersectional in design. The visibility and availability of sources impact not only how people navigate and access information, but also how they think, write, and integrate evidence in shaping their ideas and arguments. Based on our findings, we discuss the implications of these findings for designing critical engagement in AI-mediated tasks below.

\paragraph{Scaling Critical Thinking with Information Density} A central finding of our study is that the is the discovery of that effect design can reverse the negative impact of information overload on information evaluation and critical thinking. In typical search and chat interfaces, increased information density often leads to cognitive fatigue and shallower processing~\cite{buccinca2021trust, azzopardi2021cognitive, roy2022users}. It also refelects in our results: as the AI provided more citations, participants in the \textsc{Collapsible} condition (lower visibility and availability of sources) show decline in critical thinking scores when information density is high (Figure \ref{fig:interaction_plot}). The behavior patten tells the same story that designs with lower visibility and availability of sources can lead to lower engagement as conversation evolves (Figure \ref{fig:behavior_metrics_dashboard}) or less deep engagement (Table \ref{tab:sequence_transitions_top6}) . However, our \textsc{Sidebar} condition (higher visibility and availability of sources) reversed this trend: as the AI provided more citations, participants in the \textsc{Sidebar} group showed significantly improved scores in \textit{Synthesizing Multiple Sources} and \textit{Source-Related Critical Thinking}. We may attribute this to the role of the sidebar as an \textit{external working memory}. By presenting sources from the ephemeral, linear flow of the chat and placing them in a persistent, spatial layout, the \textsc{Sidebar} allows users to ``offload'' the cognitive burden of tracking evidence. Our findings echo prior work showing that transparency cues do not have uniform effects on user trust or understanding, but instead interact with task context and interface design~\cite[e.g.][]{kim2024m, kizilcec2016much, dodge2019explaining}. We also extend these insights by providing behavioral and critical-thinking evidence that transparency effects in conversational AI are strongly conditioned by information density and interaction layout rather than source presence alone.

\paragraph{The Trade-off Between Flow and Verification} Our behavioral sequence analysis reveals a distinct trade-off between interfaces that support ``flow'' and those enhance ``verification''. The \textsc{Hover} interface was unique in supporting a seamless \textit{Cite} $\to$ \textit{Write} loop (Prob. = 0.393), allowing users to verify specific claims on-demand without breaking their drafting context. This resulted in higher scores for \textit{Evaluating Evidence Strength}, as users could perform micro-verifications of individual facts. However, this efficiency came at the cost of synthesis. The \textsc{Sidebar} condition promoted a more disruptive \textit{Write} $\to$ \textit{Explore} loop, encouraging users to stop writing and scan multiple sources. While this friction likely contributed to the lower perceived usefulness ratings for the \textsc{Sidebar} if information density is low, it ultimately supported better holistic argumentation. This trade-off implies that there is no single optimal transparency design. Instead, conversational AI systems may need interfaces tailored to different tasks and information densities, or adaptive interfaces—for example, using hover-based interactions during rapid drafting to maintain flow and transitioning to persistent sidebars during review or ideation to support synthesis and complex reasoning. This trade-off resonates with prior findings that interaction friction can function as a cognitive forcing mechanism that reduces over-reliance on AI while simultaneously increasing perceived workload~\cite[e.g.][]{buccinca2021trust, bordt2022post}. This trade-off is further consistent with previous studies indicating that user trust calibration depends not only on explanation availability but also on how explanations are integrated into information interaction workflows~\cite{liao2020questioning, lai2019human}.

\paragraph{Design Safety Zone} Our findings reveal a critical ``design safety zone'' for responsible AI. A common fear in interface design is that adding friction—such as the persistent presence of a Sidebar or the demand for verification—will degrade the user experience, causing users to lose trust or feel less satisfied. However, our results show that perceptions of trust, transparency, satisfaction, and other experience metrics are resilient to interface variation. Users maintained the same level of trust in the system whether they were using the low-friction \textsc{Collapsible} interface or the high-visibility \textsc{Aligned Sidebar}. This stability provides designers with a safe zone to add friction.

\section{Conclusion}
This work revealed a critical trade-off between interface designs that prioritize workflow fluency and those that support cognitive persistence. While low-friction mechanisms like hover cards facilitate immediate, on-demand verification during drafting, persistent layouts like the aligned sidebar act as external working memory, enabling users to sustain critical synthesis even as information density increases. Consequently, there is no ``one-size-fits-all'' solution for citation display. To foster responsible human-AI collaboration, future systems must move beyond static citations toward adaptive interfaces that balance the cognitive needs of reading, verification, and synthesis based on the nature of information-seeking tasks. Future research should further analyze the prompting process of participants and explore adaptive interfaces that dynamically shift between low-friction and persistent citation layouts based on real-time detection of user intent (e.g., browsing vs. critical analysis). Additionally, longitudinal studies are needed to determine if the benefits of ``reflective friction'' persist over time.

\section{Generative AI Usage Statement}
During the preparation of this work, the authors used Gemini 3 to assist with data processing code, grammar, and latex formatting. GPT5.1 API was used to assess participants' submission data.

%\begin{acks}
%To Robert, for the bagels and explaining CMYK and color spaces.
%\end{acks}

\bibliographystyle{ACM-Reference-Format}
\bibliography{sample-base}

\appendix
\section*{APPENDIX}
\section{System Prompt}
\label{sec:system_prompt}

The following system prompt was used to configure the AI agent's behavior:

\begin{quote}
You are a research assistant designed to help users gather information for writing evidence-based essays. Your primary goal is to support critical thinking and informed decision-making.

\subsection*{Core Principles}
\begin{enumerate}
    \item \textbf{Balanced Perspectives}: When discussing controversial topics, present multiple viewpoints fairly. Help users understand different sides of complex issues.
    \item \textbf{Critical Thinking}: Encourage users to evaluate evidence, consider counterarguments, and form well-reasoned conclusions.
    \item \textbf{Encourage Exploration}: Suggest related angles or questions the user might want to explore.
    \item \textbf{Support Evidence-Based Writing}: Help users understand how to use information effectively in their essays.
\end{enumerate}

\subsection*{What NOT to Do}
\begin{itemize}
    \item Don't express personal opinions or political biases.
    \item Don't write the essay for the user — guide them in their own research and thinking.
    \item Don't make claims that go beyond what your sources support.
    \item Don't provide very lengthy responses. No more than 400 words.
\end{itemize}

Your role is to be a knowledgeable, impartial research partner who helps users develop well-informed perspectives through quality information and critical analysis.
\end{quote}

\section{Sidebar Layout Algorithm}
\label{sidebar algorithm}
We formulate the sidebar citation card placement as a bounded isotonic regression problem, which admits a globally optimal solution in linear time. For each citation card $i \in \{1, \ldots, n\}$, let $t_i$ denote its target vertical position (derived from in-text occurrences), $h_i$ its height, and $y_i$ its center position.
We minimize the weighted squared distance from targets:
$\min_{y_1, \ldots, y_n} \sum_{i=1}^{n} w_i (y_i - t_i)^2$
subject to non-overlapping constraints $y_{i+1} - y_i \geq (h_i + h_{i+1})/2 + g$ and viewport bounds $\ell_i \leq y_i \leq u_i$, where $g$ is the minimum gap between cards. Through a variable transformation, this reduces to bounded isotonic regression, solved optimally in $O(n)$ time via the Pool Adjacent Violators Algorithm (PAVA).For citations appearing multiple times, we compute target positions as weighted averages with exponential decay $t_i = \sum_{j} e^{-0.6(j-1)} p_{i,j} / \sum_{j} e^{-0.6(j-1)}$, where $p_{i,j}$ are occurrence positions in reading order.
This prioritizes alignment with first mentions. To maintain strict alignment during interaction, we dynamically adjust $w_i$ during scrolling based on proximity to the viewport center, applying a decay factor $\max(0.1, 1 - d_i/D)$ where $d_i$ is the distance from the center and $D$ is the maximum viewport extent. This ensures visible citations track their targets tightly while off-screen cards absorb necessary displacements.
\section{Instructional Videos}
\label{sec:instruction_videos}

The instructional videos for each interface condition can be accessed via the following links:

\begin{itemize}
    \item \textsc{Collapsible} Interface: \url{https://youtu.be/xyxuJ9L3QG4}
    \item \textsc{Hover Card} Interface: \url{https://youtu.be/l9zDlvxMAKU}
    \item \textsc{Footer} Interface: \url{https://youtu.be/ziP2O9-sYLY}
    \item \textsc{Aligned Sidebar} Interface: \url{https://youtu.be/60K3foLeqe8}
\end{itemize}

\section{Regression Analysis}
\label{sec:regression_analysis}

\begin{table}[htbp]
    \centering
    \begin{threeparttable}
    \caption{Summary of Adjusted Regression Models for Perception and Experience Outcomes}
    \label{tab:perception_results_complete}
    \small
    \setlength{\tabcolsep}{3pt}
    \begin{tabular}{lccccccc}
        \toprule
        \textbf{Predictor} & \textbf{Transparency} & \textbf{Trust} & \textbf{Confidence} & \textbf{Satisfaction} & \textbf{Usefulness} & \textbf{Quality} & \textbf{Workload} \\
        \midrule
        \multicolumn{8}{l}{\textit{UI Conditions (ref: \textsc{Collapsible})}} \\
        UI: \textsc{Sidebar}             & -0.37    & -0.44    & -0.62    & -1.26    & -1.41*   & -0.43    & -0.66    \\
        UI: \textsc{Hover}               & 1.15     & 0.24     & 0.29     & 0.25     & -0.18    & -0.13    & -1.12    \\
        UI: \textsc{Footer}              & 0.38     & -0.33    & 0.87     & 0.27     & -0.24    & -0.06    & -0.37    \\
        \midrule
        \multicolumn{8}{l}{\textit{Interactions}} \\
        \textsc{Sidebar} $\times$ Citations & 0.03     & 0.02     & 0.03     & 0.05     & 0.04*    & 0.01     & 0.03     \\
        \textsc{Hover} $\times$ Citations& -0.03    & -0.01    & 0.00     & -0.01    & -0.01    & 0.00     & 0.05*    \\
        \textsc{Footer} $\times$ Citations  & -0.01    & 0.02     & -0.03    & 0.00     & 0.01     & -0.00    & 0.02     \\
        \midrule
        \multicolumn{8}{l}{\textit{Controls}} \\
        Age                     & -0.00    & -0.00    & -0.01    & 0.01     & 0.00     & -0.00    & 0.02**   \\
        Sex Female              & 0.42*    & -0.03    & -0.16    & 0.24     & 0.19     & 0.47*    & 0.16     \\
        Education               & 0.04     & -0.13    & -0.02    & 0.03     & 0.07     & 0.07     & 0.17     \\
        Llm\_usage\_freq        & -0.06    & -0.19**  & -0.09    & -0.16*   & -0.19**  & -0.16*   & -0.10    \\
        Search\_usage\_freq     & -0.17    & 0.28     & -0.07    & -0.11    & 0.20     & 0.22     & 0.05     \\
        Eval\_rel\_confidence   & 0.97***  & 0.80***  & 0.88***  & 0.90***  & 0.78***  & 0.59**   & -0.16    \\
        Eval\_trus\_confidence  & -0.06    & 0.01     & 0.11     & -0.15    & 0.18     & 0.17     & -0.06    \\
        Knowledge\_pre          & -0.14    & -0.10    & 0.23**   & -0.08    & -0.10    & -0.03    & -0.12    \\
        Interest\_pre           & 0.06     & -0.01    & 0.02     & 0.09     & 0.02     & 0.02     & 0.18*    \\
        Agreement\_pre          & 0.11     & 0.16*    & 0.06     & -0.00    & 0.07     & 0.11     & -0.01    \\
        Humility                & 0.41**   & 0.36**   & -0.03    & 0.24     & 0.32*    & 0.27     & 0.02     \\
        \midrule
        $R^2$ (Pseudo)          & 0.09     & 0.07     & 0.09     & 0.08     & 0.09     & 0.06     & 0.03     \\
        N                       & 372      & 372      & 372      & 372      & 372      & 372      & 372      \\
        \bottomrule
    \end{tabular}
    \begin{tablenotes}
        \item \textit{Note:} * $p<0.05$, ** $p<0.01$, *** $p<0.001$.
    \end{tablenotes}
    \end{threeparttable}
\end{table}

\begin{table}[htbp]
    \centering
    \begin{threeparttable}
    \caption{Summary of Adjusted Regression Models (Model 2) for All Critical Thinking Outcomes}
    \label{tab:model2_summary_complete}
    \small
    \setlength{\tabcolsep}{3pt}
    \begin{tabular}{lcccccccc}
        \toprule
        \textbf{Predictor} & \textbf{Synthes.} & \textbf{Eval.} & \textbf{Counterarg.} & \textbf{Facts/Opinions} & \textbf{Concl.} & \textbf{Fallacies} & \textbf{Source-Rel.} & \textbf{Overall} \\
        \midrule
        \multicolumn{9}{l}{\textit{UI Conditions (ref: \textsc{Collapsible})}} \\
        UI: \textsc{Sidebar}             & -1.53*   & -0.55    & -0.86    & -1.09    & -1.05    & -1.78*   & -0.36    & -0.36*  \\
        UI: \textsc{Hover}               & -1.16    & -1.48    & -0.18    & -0.90    & -0.25    & -0.68    & -0.42    & -0.24   \\
        UI: \textsc{Footer}              & -0.95    & 0.15     & -0.45    & -0.76    & -1.52    & -0.72    & -0.27    & -0.25   \\
        Citations               & -0.02    & -0.01    & 0.00     & -0.02    & -0.02    & -0.04*   & -0.01    & -0.01   \\
        \midrule
        \multicolumn{9}{l}{\textit{Interactions}} \\
        \textsc{Sidebar} $\times$ Citations & 0.07**   & 0.05     & 0.02     & 0.07**   & 0.04     & 0.07**   & 0.02**   & 0.02**  \\
        \textsc{Hover} $\times$ Citations& 0.06     & 0.08*    & -0.00    & 0.05     & 0.02     & 0.04     & 0.02*    & 0.01    \\
        \textsc{Footer} $\times$ Citations  & 0.04     & 0.01     & 0.01     & 0.05     & 0.06     & 0.03     & 0.01     & 0.01    \\
        \midrule
        \multicolumn{9}{l}{\textit{Controls}} \\
        Age                     & -0.03*** & -0.03*** & -0.01    & -0.03*** & -0.02**  & -0.03*** & -0.01*** & -0.01***\\
        Sex Female              & 0.35     & 0.42     & 0.16     & 0.46*    & 0.14     & 0.47*    & 0.16*    & 0.12*   \\
        Education               & 0.49***  & 0.28*    & 0.04     & 0.39***  & 0.16     & 0.36**   & 0.13***  & 0.09**  \\
        Llm\_usage\_freq        & 0.09     & 0.01     & 0.04     & 0.07     & 0.04     & 0.02     & 0.02     & 0.02    \\
        Search\_usage\_freq     & 0.06     & 0.08     & -0.36*   & -0.11    & -0.11    & -0.09    & 0.01     & -0.04   \\
        Eval\_rel\_confidence   & 0.05     & 0.32     & 0.15     & 0.08     & 0.11     & 0.10     & 0.05     & 0.05    \\
        Eval\_trus\_confidence  & 0.27     & 0.06     & 0.03     & 0.31     & 0.08     & 0.14     & 0.07     & 0.05    \\
        Knowledge\_pre          & -0.18*   & -0.28**  & -0.22**  & -0.16    & -0.29**  & -0.21*   & -0.06*   & -0.07***\\
        Interest\_pre           & 0.18*    & 0.16     & 0.04     & 0.06     & 0.09     & 0.12     & 0.03     & 0.03    \\
        Agreement\_pre          & 0.08     & 0.18*    & 0.08     & 0.14     & 0.06     & 0.17*    & 0.05     & 0.04*   \\
        Humility                & 0.19     & 0.16     & 0.15     & 0.16     & 0.22     & 0.52***  & 0.09     & 0.08*   \\
        \midrule
        $R^2$                   & 0.07     & 0.09     & 0.03     & 0.07     & 0.05     & 0.10     & 0.17     & 0.18    \\
        N                       & 371      & 371      & 371      & 371      & 371      & 371      & 365      & 371     \\
        \bottomrule
    \end{tabular}
    \begin{tablenotes}
        \item \textit{Note:} * $p<0.05$, ** $p<0.01$, *** $p<0.001$. Model 2 adjusted for full set of 18 predictors.
    \end{tablenotes}
    \end{threeparttable}
\end{table}
\newpage
\section{Survey Instruments}
\label{sec:survey_instruments}

\subsection{Pre-Task Survey}

\subsubsection{Demographics and Experience}
\begin{enumerate}
    \item \textbf{What is the highest level of education you have completed?}
    \begin{itemize}
        \item High school
        \item Some college
        \item Bachelor's degree
        \item Master's degree
        \item Doctorate
        \item Other
    \end{itemize}

    \item \textbf{How often do you use large language models (LLMs) such as ChatGPT, Claude, Gemini, Grok, etc.?}
    \begin{itemize}
        \item Daily
        \item 4-6 times a week
        \item 2-3 times a week
        \item Once a week
        \item 2-3 times per month
        \item Less than 2-3 times per month or never
    \end{itemize}

    \item \textbf{For what types of tasks do you generally use large language models?} (Select all that apply)
    \begin{itemize}
        \item Writing
        \item Learning/education
        \item Programming
        \item Information search/retrieval
        \item Work-related tasks
        \item Entertainment
        \item Therapy
        \item Other
    \end{itemize}

    \item \textbf{How often do you use Web search engines (e.g. Google, Microsoft Bing, Yahoo! Search) for information searching and evaluation?}
    \begin{itemize}
        \item Daily
        \item 4-6 times a week
        \item 2-3 times a week
        \item Once a week
        \item Less than once a week
    \end{itemize}
\end{enumerate}

\subsubsection{Topic Assessment}
\begin{enumerate}
    \item \textbf{How much do you know about the topic [Assigned Topic]?}
    \begin{itemize}
        \item 7-point scale: Not knowledgeable at all -- Extremely knowledgeable
    \end{itemize}

    \item \textbf{How interested you are in the topic [Assigned Topic]?}
    \begin{itemize}
        \item 7-point scale: Not at all interested -- Extremely interested
    \end{itemize}

    \item \textbf{To what extent do you agree or disagree with the following statement? [Assigned Question Statement]}
    \begin{itemize}
        \item 7-point scale: Strongly disagree -- Strongly agree
    \end{itemize}
\end{enumerate}

\subsection{Post-Task Survey}
\textit{Participants rated \textit{Transparency} and \textit{Trust} using 3-item scales on 7-point Likert scales (1=Strongly Disagree, 7=Strongly Agree). The measures were adopted from Shin \cite{shin2021effects}.}
\subsubsection{Transparency and Trust}
\textbf{Please indicate your level of agreement with the following statements regarding the sources and references.} (7-point scale: Strongly disagree -- Strongly agree)
\begin{itemize}
    \item The way the chatbot selects and presents sources for my essay topic is understandable to me.
    \item The chatbot clearly indicates where specific information comes from, making it easy to verify the original text.
    \item The chatbot allows me to see the connection between the cited reference and the text it wrote for me.
\end{itemize}

\textbf{Please indicate your level of agreement with the following statements regarding the utility of the responses.} (7-point scale: Strongly disagree -- Strongly agree)
\begin{itemize}
    \item I would trust the information provided by this chatbot to be included in my actual essay.
    \item I believe that the references and citations generated by the chatbot are trustworthy.
    \item I believe that the writing assistance provided by the chatbot is reliable.
\end{itemize}

\subsubsection{User Experience}
\begin{enumerate}
    \item \textbf{To what extent are you confident with the quality of your essay submission?}
    \begin{itemize}
        \item 5-point scale: Very unconfident -- Very confident
    \end{itemize}

    \item \textbf{To what extent are you satisfied with your chat interaction experience?}
    \begin{itemize}
        \item 5-point scale: Very unsatisfied -- Very satisfied
    \end{itemize}

    \item \textbf{To what extent do you feel that the information sources/links provided in the chat response are useful?}
    \begin{itemize}
        \item 5-point scale: Not at all useful -- Extremely useful/perfect
    \end{itemize}

    \item \textbf{To what extent do you feel that the overall quality of responses meet your prior expectations on the LLM chatbot?}
    \begin{itemize}
        \item 5-point scale: Far below expectations -- Far above expectations
    \end{itemize}
\end{enumerate}

\subsubsection{Workload (NASA-TLX)}
Please respond to each item on a 7-point scale (Very low -- Very high):
\begin{itemize}
    \item How mentally demanding was the task?
    \item How hurried or rushed was the pace of the task?
    \item How successful were you in accomplishing what you were asked to do?
    \item How hard did you have to work to accomplish your level of performance?
\end{itemize}

\subsubsection{Topic Re-assessment}
\textit{Participants re-evaluated their knowledge, interest, and agreement with the topic using the same scales as the Pre-Task Survey.}

\subsubsection{Intellectual Humility and Confidence}
\textbf{Please respond to each item on a 5-point scale (Does not describe me -- Describes me extremely well):}
\begin{itemize}
    \item I question my own opinions, positions, and viewpoints because they could be wrong.
    \item I reconsider my opinions when presented with new evidence.
    \item I recognize the value in opinions that are different from my own.
    \item I like finding out new information that differs from what I already think is true.
    \item I accept that my beliefs and attitudes may be wrong.
    \item In the face of conflicting evidence, I am open to changing my opinions.
\end{itemize}

\textbf{Information Evaluation Confidence:} (5-point scale: Very unconfident -- Very confident)
\begin{itemize}
    \item How confident are you in evaluating the topical relevance of online information to your search queries?
    \item How confident are you in evaluating the trustworthiness of online information?
\end{itemize}

\end{document}